\documentclass[12pt]{article}

\catcode`\@=11

\global\arraycolsep=2pt
\usepackage{amsbsy,amssymb,latexsym,amsfonts,amsmath}
\usepackage{graphicx,color}

\begin{document}
\begin{flushright}
\parbox{4.2cm}
{RUP-17-23}
\end{flushright}

\vspace*{0.7cm}

\begin{center}
{ \Large Cancelling Weyl anomaly from  position dependent coupling}
\vspace*{1.5cm}\\
{Yu Nakayama}
\end{center}
\vspace*{1.0cm}
\begin{center}

Department of Physics, Rikkyo University, Toshima, Tokyo 171-8501, Japan

\vspace{3.8cm}
\end{center}

\begin{abstract}
Once we put a quantum field theory on a curved manifold, it is natural to further assume that coupling constants are position dependent. The position dependent coupling constants then provide an extra contribution to the Weyl anomaly so that we may attempt to cancel the entire Weyl anomaly on the curved manifold. We show that such a cancellation is possible for constant Weyl transformation or infinitesimal but generic Weyl transformation in two and four dimensional conformal field theories with exactly marginal deformations. When the Weyl scaling factor is annihilated by conformal powers of Laplacian (e.g. by Fradkin-Tseytlin-Riegert-Paneitz operator in four dimensions), the cancellation persists even at the finite order thanks to a nice mathematical property of the $Q$-curvature under the Weyl transformation.
\end{abstract}

\thispagestyle{empty} 

\setcounter{page}{0}

\newpage




\section{Introduction}
The Weyl anomaly in quantum field theory has a long history (see e.g. \cite{Duff:1993wm} for a historical overview). It states that the Weyl transformation of the metric in the curved space-time may not be a symmetry of the system even though the quantum field theory under consideration has the conformal symmetry in the flat space-time. Indeed, in most conformal field theories, the Weyl anomaly is non-vanishing, and we say that the Weyl symmetry is quantum mechanically broken in curved space-time.

The Weyl anomaly has played many important roles in theoretical physics.
The existence of the Weyl anomaly gives a constraint on the expectation values of energy-momentum tensor in curved background \cite{Page:1982fm}, and may be related to the nature of Hawking radiation \cite{Robinson:2005pd}\cite{Iso:2006wa}\cite{Kawai:2017txu}\cite{Kawai:2014afa}.  The fact that vanishing of the Weyl anomaly happens only in a limited class of theories dictates the number of space-time dimensions in critical string theory. 
More recently, we find that the universal terms in the entanglement entropy of conformal field theories are given by the coefficient of the Weyl anomaly, suggesting a deep relation between geometry and information \cite{Ryu:2006ef}.

What we would like to study in this paper is to find a way to cancel the Weyl anomaly from the other source, e.g. from the position dependent coupling constant.\footnote{Sorry for the oxymoron. It is no longer constant. Probably it is Dirac who openly advocated this idea in the early days \cite{Dirac:1938mt}.} Once we put a quantum field theory on a curved manifold, it is natural to assume that coupling constants are position dependent. The position dependent coupling constants then provide an extra contribution to the Weyl anomaly so that we may attempt to cancel the entire Weyl anomaly on the curved manifold. We would like to find under which condition such a cancellation is possible.

The similar idea of cancelling more general anomalies have been implicitly assumed in many places. For example, if we try to introduce background gauge fields for chiral current operators in the curved background (e.g. in the context of supersymmetric localization), then they may be mutually inconsistent due to the 't Hooft anomaly. One way to avoid this is to cancel the anomaly of the background gauge field from the space-time curvature and vice versa. Similarly, if preserving the Weyl anomaly is the critical issue (e.g. if we try to gauge it), our new way of doing it may be another option to be considered.

The organization of the paper is as follows. In section 2, we study the cancellation of the Weyl anomaly from position dependent coupling constant in two dimensional conformal field theories. In section 3, we study it in three dimensions and in section 4, we study it in four dimensions. We supplement the holographic viewpoint in section 5 and conclude with some discussions in section 6.

\section{Two dimensions}
Let us consider a two-dimensional conformal field theory with an exactly marginal deformation denoted by $g$. For instance, we may take the Gaussian $c=1$ boson with the compactification radius as the exactly marginal deformation here.
We put the theory on a curved background with the metric $g_{\mu\nu}(x)$ and then vary the coupling constant $g(x)$ over the manifold: schematically we consider the action $S = S_0 + \int d^2x\sqrt{g} g(x) O(x)$. Even though the theory is conformal invariant in the Euclidean space $g_{\mu\nu} = \delta_{\mu\nu}$ with $g(x) = g$, it is not necessarily so after turning on the background metric and position dependent coupling constant.\footnote{We define the scale transformation  by the change of the difference of the coordinate as in \cite{Osborn:1991gm} rather than the coordinate itself \cite{Dong:2012ua}.} 
This obstruction is known as the Weyl anomaly under the infinitesimal Weyl rescaling: $\delta g_{\mu\nu}(x) = 2\delta \sigma g_{\mu\nu}$. 

In terms of the free energy functional $e^{-F[g_{\mu\nu}(x),g(x)]} = \int \mathcal{D} \Phi e^{-S[\Phi]}$, the Weyl anomaly for a two-dimensional conformal field theory is given by (e.g. see  \cite{Osborn:1991gm})\footnote{In this paper, we always assume that the conformal field theories under consideration preserve the CP symmetry.}
\begin{align}
\delta F_{\sigma} = \int d^2x\sqrt{g} \delta \sigma(x) (cR -\frac{1}{2} \partial^\mu g(x) \partial_\mu g(x)) \  \label{twoa}
\end{align}
in a certain renormalization scheme so that the exactly marginal deformation has a flat line metric. Otherwise, we can always redefine the coupling constant or the renormalization scheme so that it is flat. It is clear that when $g(x) = g$, the only way to cancel the Weyl anomaly is to require $c=0$, which is typically what we demand in critical string theory.

However, we see that this is not the only available option. Now, given a positive curvature $R(x)\ge 0$, we may try to cancel the curvature term in the Weyl anomaly \eqref{twoa} against the second term originating from the position dependent coupling constant by solving the equation
\begin{align}
c R = \frac{1}{2} \partial^\mu g(x) \partial_\mu g(x) \ . \label{twoc}
\end{align}
This is possible for positive curvature $R(x) \ge 0$ (assuming $c>0$ in unitary conformal field theories).

For example, if we take the Fubini-Study metric on the sphere with the complex coordinate $z$ and $\bar{z}$:
\begin{align}
ds^2 = \frac{dzd\bar{z}}{(1+|z|^2)^2} \ , 
\end{align}
the solution of \eqref{twoc} is 
\begin{align}
g(x) = \sqrt{c} \cdot \mathrm{arctan}(|z|)  + \mathrm{const} \ .
\end{align}
In this way, one can cancel the Weyl anomaly on the sphere by introducing the position dependent coupling constant. Note, however, that the position dependence of the coupling constant reduces the symmetry of the sphere from $SO(3)$ down to $SO(2)$. The idea here is we gained extra ``Weyl symmetry" at the sacrifice of the rotational symmetry.\footnote{This is not necessarily a bad idea: for examle, in the supersymmetric localization, we often do not keep the full isometry of the sphere but only the $U(1)$ subgroup of it.}

The above cancellation works both for infinitesimal generic Weyl transformation $\delta g_{\mu\nu}(x) = 2\delta \sigma(x) g_{\mu\nu}(x)$ or finite but constant Weyl transformation $g_{\mu\nu}(x) \to e^{2\bar{\sigma}} g_{\mu\nu}(x)$, where $\bar{\sigma}$ is a finite constant. The latter is because the equation to be solved in \eqref{twoc} trivially scales under the constant Weyl transformation, so once it is solved then it is also solved after finite but constant Weyl transformation.
For finite generic Weyl transformation, however, the cancellation may not persist. The point is that the curvature term in the Weyl anomaly is non-trivially transforms under the Weyl transformation:
\begin{align}
R \to e^{-2\sigma(x)} (R -2D^2 \sigma) \ , 
\end{align}
where $D^2$ is the Laplacian, 
while the Weyl transformation of the second term from the position dependent coupling constant $\partial^\mu g(x) \partial_\mu g(x)$ is trivial:
\begin{align}
\partial^\mu g(x) \partial_\mu g(x) \to e^{-2\sigma(x)} \partial^\mu g(x) \partial_\mu g(x)
\end{align}
Thus, even though one may solve the cancellation condition for a given $g_{\mu\nu}(x)$ with a certain position dependent coupling constant $g(x)$, the cancellation does not persist for the Weyl transformed geometry. 

 Nevertheless, we realize that the cancellation is still intact if we restrict\footnote{A similar restriction on the Weyl transformation has been studied in \cite{Edery:2014nha}.} ourselves to the harmonic Weyl transformation, which satisfies $D^2 \sigma = 0$. Thus, we may construct a quantum field theory which is exactly invariant under the harmonic Weyl transformation by cancelling the Weyl anomaly from the position dependent coupling constant.

More generically, one may consider theories with several exactly marginal deformations. The Weyl anomaly has the generalized form
\begin{align}
\delta F_{\sigma} = \int d^2x\sqrt{g} \delta\sigma(x) (cR - \chi_{ij}(g) \partial^\mu g^i(x) \partial_\mu g^j(x)) + \partial_\mu \delta\sigma(x) w_i(g) \partial^\mu g^i(x)  \ , \label{twog}
\end{align}
where $\chi_{ij}(g)$ and $w_i(g)$ may depend on the exactly marginal deformations $g^i(x)$. 
For a constant Weyl transformation, the condition for the cancellation is essentially the same as before since the last term in \eqref{twog} drops out.

For infinitesimal generic Weyl transformation, however, we have to think about the cancellation of the third term proportional to $\partial_\mu \delta \sigma(x)$. We did not talk about it in the single coupling case because we were able to remove it from the local counterterm $\int d^2x\sqrt{g} b(g) R$, but we have to discuss it now with several coupling constants when it has the non-trivial curvature $\partial_i w_j - \partial_j w_i$.
While the Wess-Zumino consistency condition does not say anything about the (non-)existence of this term  \cite{Osborn:1991gm}, the recent analysis in \cite{Gomis:2015yaa} tells that on the conformal manifold spanned by the exactly marginal deformations, the curvature is trivial (i.e.  $\partial_i w_j - \partial_j w_i = 0$) and can be removed by the local counterterm $\int d^2x\sqrt{g} b(g^i) R$,
 so we actually do not have to worry about its cancellation. The non-existence of the curvature $\partial_i w_j -\partial_j w_i$ is related to the gradientness of the beta functions and it may have a deep implication in renormalization group flows  \cite{Osborn:1991gm}\cite{Friedan:2009ik}\cite{Gukov:2016tnp}.

\section{Three dimensions}
There is no curvature dependent Weyl anomaly in three dimensions. The position dependent exactly marginal deformations do not introduce the additional Weyl anomaly either under the assumption of the CP symmetry \cite{Nakayama:2013wda}. Thus there is no interesting scenario we can imagine in three dimensions.

\section{Four dimensions}
Let us consider a four-dimensional conformal field theory with an exactly marginal deformation denoted by $g$. We put the theory on a curved background with the metric $g_{\mu\nu}(x)$ and then vary the coupling constant over the manifold $g(x)$. 
In a certain renormalizaiton scheme, the first order Weyl transformation (i.e. the Weyl anomaly) is given by \cite{Osborn:1991gm} (See also \cite{Nakayama:2013is}\cite{Jack:2013sha}.)
\begin{align}
\delta F_{\sigma}  = \int d^4x \sqrt{g} \delta\sigma(x) &\left( c(g) \mathrm{Weyl}^2 - a \mathrm{Euler} + (D^2 g D^2g - 2G_{\mu\nu} \partial^\mu g \partial^\nu g -\frac{R}{3}\partial^\mu g\partial_\mu g)  \right. \cr
& \left. + \chi_4(g) \partial_\mu g \partial^\mu g \partial_\nu g \partial^\nu g \right) \ .  \label{foura}
\end{align}
Here $G_{\mu\nu} = R_{\mu\nu} - \frac{R g_{\mu\nu}}{2}$ is the Einstein tensor and $D^\mu$ is the covariant derivative. In addition, we have introduced $\mathrm{Weyl}^2 = R_{\mu\nu\rho\sigma}^2 -2R_{\mu\nu}^2 + \frac{1}{3}R^2$ and $\mathrm{Euler} = R_{\mu\nu\rho\sigma}^2 - 4R_{\mu\nu}^2 + R^2$. 
In principle $c(g)$ can depend on $g$, but the only such theories known are constructed in somewhat artificial holographic realization \cite{Nakayama:2017oye}.

To simplify the analysis, let us focus on the regime in which the last quartic term in \eqref{foura} i.e. $\chi_4(g) \partial_\mu g \partial^\mu g \partial_\nu g \partial^\nu g$ can be neglected (e.g. in the small coupling regime). 
Neglecting the quartic term, we try to solve the equation
\begin{align}
-c\mathrm{Weyl}^2 + a \mathrm{Euler} = (D^2 g D^2g - 2G_{\mu\nu} \partial^\mu g \partial^\nu g -\frac{R}{3}\partial^\mu g\partial_\mu g) \ . \label{fourc}
\end{align}
In particular, suppose that the metric $g_{\mu\nu}(x)$ is Ricci flat. Then the equation \eqref{fourc} becomes
\begin{align}
\sqrt{(a-c) R_{\mu\nu\rho\sigma}^2} = D^2 g \ , 
\end{align}
which may be solved by using Green's function for the Laplacian 
\begin{align}
g(x) = \int d^4x' G(x,x') \sqrt{(a-c) R_{\mu\nu\rho\sigma}^2(x')}
\end{align}
when the manifold is non-compact (otherwise the regular solution does not exist).

As in two-dimensions, the above argument works both for infinitesimal Weyl transformation or finite but constant Weyl transformation.
For finite generic Weyl transformation, one may define the analogue of harmonic Weyl transformation. For this purpose, it is more convenient to choose a different renormalization scheme so that the Weyl anomaly takes the form (e.g. \cite{Gomis:2015yaa})
\begin{align}
\delta F_{\sigma}  = \int d^4x \sqrt{g} \delta\sigma(x) &\left( (c(g)-a) \mathrm{Weyl}^2 - 4a Q   + g \Delta_4 g \right. \cr
& \left. + \chi_4(g) \partial_\mu g \partial^\mu g \partial_\nu g \partial^\nu g \right) \ , \label{alt} 
\end{align}
where $\Delta_4$ is the Fradkin-Tseytlin-Riegert-Paneitz conformal operator \cite{Fradkin:1982xc}\cite{Fradkin:1981jc}\cite{Riegert:1984kt}\cite{PA}
\begin{align}
\Delta_4 = (D^2)^2 + 2G_{\mu\nu}D^\mu D^\nu + \frac{1}{3}(D^\mu R) D_\mu + \frac{1}{3}R D^2, 
\end{align}
which is Weyl covariant $\Delta_4 \to e^{-4\sigma} \Delta_4$, 
and $Q$ is what is called the Q-curvature \cite{Q}:
\begin{align}
Q = \frac{-1}{6} D^2 R -\frac{1}{2}R^{\mu\nu}R_{\mu\nu} + \frac{1}{6}R^2 \ 
\end{align}
which has a nice mathematical property under the Weyl transformation
\begin{align}
Q \to e^{-4\sigma} (Q + \Delta_4 \sigma)
\end{align}

The advantage of this rewriting or a choice of the particular local counterterm is as follows. Suppose we cancelled the Weyl anomaly at a particular background by demanding
\begin{align}
0 = (c(g)-a) \mathrm{Weyl}^2 - 4a Q   + g \Delta_4 g + \chi_4(g) \partial_\mu g \partial^\mu g \partial_\nu g \partial^\nu g 
\end{align}
Then, we are still able to cancel the Weyl anomaly on the Weyl transformed manifold whenever the Weyl rescaling is annihilated by the Fradkin-Tseytlin-Riegert-Paneitz operator:
\begin{align}
\Delta_4 \sigma = 0 \ . \label{ann}
\end{align}
This is because all the terms in \eqref{alt} except for the Q-curvature transform covariantly under the finite Weyl transformation.
If the Weyl scaling factor satisfies \eqref{ann}, the cancellation of the Weyl anomaly therefore persists even for finite Weyl transformation. This is precisely analogous to the special role of harmonic Weyl transformation in two dimensions.

Let us move on to the most generic cases with multiple coupling constants. The Weyl transformation is given by
\begin{align}
\delta F_{\sigma}  = \int d^4x \sqrt{g} \delta \sigma(x) &( c(g) \mathrm{Weyl}^2 - a \mathrm{Euler} +  \chi_{ij}(g) (D^2 g^i D^2g^j - 2G_{\mu\nu} \partial^\mu g^i \partial^\nu g^j -\frac{R}{3}\partial^\mu g^i \partial_\nu g^j)  \cr
& \left. + \chi_{ijkl}(g) \partial_\mu g^i \partial^\mu g^j \partial_\nu g^k \partial^\nu g^l \right)  \cr
&+ \partial_\mu \delta\sigma G^{\mu\nu} w_i(g) \partial_\nu g^j \ . \label{fourm}
\end{align}
For finite but constant Weyl transformation, we only have to cancel the first two lines in \eqref{fourm}, which is  essentially equivalent to what we have done in the above. On the other hand, for infinitesimal but generic Weyl transformation, we have to cancel the third line as well, which requires either $\partial_i w_j - \partial_j w_i =0$ or $G_{\mu\nu} = 0$ in the background.

To conclude the analysis, we would like to mention the other obstructions to the Weyl transformation if the dimension two operator $O(x)$ exist in the theory. If this is the case, there is a further operator Weyl anomaly such as 
\begin{align}
\int d^4x \sqrt{g} \delta\sigma(x) (\eta(g) R O(x) + \epsilon(g) \partial_\mu g \partial^\mu g O + \tau(g)  D^2 O + \delta(g) D^2 g O) + \partial_\mu \delta \sigma(x) \theta(g) \partial^\mu g O \ .
\end{align}
It has been shown that such Weyl anomaly can be removed when $g(x) = g$ \cite{Jack:2013sha} (see also \cite{Nakayama:2013wda}\cite{Farnsworth:2017tbz} for similar analysis), but with the space-time dependent coupling, we need the extra cancellation to get the consistent picture. Schematically, the Wess-Zumino consistency condition demands $\eta = 0$ and  one can always remove $\theta$ and $\tau$ by local counterterms. Then we need to cancel $\epsilon$ term against the $\delta$ term. Since the existence of dimension two operator is non-generic, we will not pursue the cancellation in further details.

\section{Holographic models}
We revisit the cancellation mechanism we have studied in previous sections from  the holographic perspective. For definiteness we consider the case of four dimensional conformal field theories with the five dimensional bulk. Let us study the Einstein gravity coupled with a scalar field $\phi$ given by the minimal action
\begin{align}
S = \int d^5x \sqrt{g} \left( R + \Lambda + \frac{1}{2}\partial^M \phi \partial_M \phi  \right) .
\end{align}

In the AdS/CFT correspondence, we compute the on-shell action for a given boundary condition at $\rho= \epsilon$ (i.e. $\phi_{(0)}(x)$ and $g_{(0)\mu\nu}(x)$ below) with the expansion 
\begin{align}
\phi &= \phi_{(0)}(x) + \rho \phi_{(1)}(x) + \rho^2 \phi_{(2)}(x) + \cdots \cr
g_{\mu\nu} &= g_{(0)\mu\nu}(x) + \rho g_{(1)\mu\nu}(x) + \rho^2 g_{(2) \mu\nu}(x) + \cdots
\end{align}
in the Graham-Fefferman gauge 
\begin{align}
ds^2 = G_{MN} dx^M dx^N = \frac{d\rho^2}{\rho^2} +  \frac{g_{\mu\nu} dx^\mu dx^\nu}{\rho}  \ .
\end{align}

The resulting on-shell action is generically divergent in the limit $\epsilon \to 0$ from the $\rho$ integration as $\int_{\epsilon} d\rho \rho^{-1} S_{\log} = \log \epsilon S_{\log}$, leading to the holographic Weyl anomaly \cite{Henningson:1998gx}. Explicitly \cite{Nojiri:1998dh}, we have
\begin{align}
S = \log \epsilon \int d^4x &\left( \frac{1}{8} R_{\mu\nu (0)}^2 -\frac{1}{24} R_{(0)}^2  + \frac{1}{4} (D^2 \phi_{(0)})^2 \right. \cr 
& \left.  - \frac{1}{2} R_{(0)}^{\mu\nu} \partial_\mu \phi \partial_\nu \phi + \frac{1}{6} R_{(0)} \partial^\mu \phi_{(0)} \partial_\nu \phi_{(0)} + \frac{1}{3}(\partial_\mu \phi_{(0)} \partial^\mu\phi_{(0)})^2  \right) \ 
\end{align}
which is exactly what we had in section 4 for the constant Weyl transformation.
Thus cancelling the Weyl anomaly from the position dependent coupling constant corresponds to the choice of the boundary values of $\phi(x)$ such that the on-shell gravity action is finite without the logarithmic divergence. The choice of such boundary conditions make the AdS/CFT correlation functions better behaved, so classifying such supergravity background may be of theoretical interest.

\section{Discussions}
In this paper, we have studied a novel way to cancel the Weyl anomaly from the position dependent coupling constant. Here we would like to mention further possibilities to cancel the Weyl anomaly.

First of all, if the theory under consideration possesses a conserved current $J_\mu$, one may introduce the background field strength by the coupling $\int d^4x \sqrt{g} A^\mu J_\mu$. This gives another contribution to the Weyl anomaly as $\int d^4x \delta\sigma(x) b_0 F_{\mu\nu}(x) F^{\mu\nu}(x)$, where $b_0$ is the coefficient of the one-loop beta function determined from the current two-point function (which is positive in unitary conformal field theories) and $F_{\mu\nu} = \partial_\mu A_\nu - \partial_\nu A_\mu$. Then we may try to cancel the Weyl anomaly from this contribution.
Actually, simultaneous use of the gauge field and the position dependent coupling constant may not be a good idea because of the existence of the vector beta functions \cite{Nakayama:2013ssa}. Again we have to think about the cancellation of the extra operator Weyl anomaly such as $\int d^4x \delta \sigma \rho(g) \partial^\mu g J_\mu$. To avoid the appearance of the vector beta functions, we may only introduce the position dependent coupling constant which is neutral under the symmetry generated by $J_\mu$.

It could have been extremely interesting if we were able to find a novel class of Weyl gauging without demanding $c=0$ in two dimensions, and $c=a=0$ in four dimensions.\footnote{See e.g. \cite{Fradkin:1983tg} for a possibility in the context of supersymmetric Weyl gravity.} 
Currently, the closest way to do this is to demand all the non-trivial Weyl anomalies vanish, say $a=0$ in four dimensions, and then try to cancel the $c \mathrm{Weyl}^2$ term against the space-time dependent coupling constants. Here, we should further employ the non-unitariness of the model (since $a=0$ from the beginning suggests it must be so) to obtain the cancellation in the Weyl anomalies. This is because unitarity demands the positivity of the both terms and the cancellation only happens by using the non-unitary property. Whether this is better than just demanding $c=a=0$ is yet to be seen in the context of quantum Weyl gravity in which we would like to gauge the Weyl symmetry exactly.\footnote{Note, however, that the introduction of the position dependent coupling constant modifies the conservation of the energy-momentum tensor, and the consistecy with the dynamical gravity must be discussed more carefully. We stress that in the main part of this paper, we have focused on the fixed gravitational background, so there is no inconsistency. The author would like to thank S.~Deser for the discussions.}

In two dimensions, we did not obtain any new possibilities to gauge the entire Weyl symmetry than demanding $c=0$ from the beginning. We are still able to gauge the harmonic Weyl symmetry, but the physical interest in such gauging (e.g. whether it defines new class of quantum gravity in two dimensions) should be discussed more in detail.

Finally, we point out that there is an alternative option. Once we know how to solve $g(x)$ to cancel the Weyl anomaly in a given metric $g_{\mu\nu}(x)$, one may introduce the extra transformation on $g(x)$ so that the Weyl anomaly is always cancelled (irrespective of the obstructions we have discussed above). This possibility requires further investigation if such transformation can be defined systematically and then whether such a generalized notion of the Weyl transformation is useful or not.


\section*{Acknowledgements}
This work is in part supported by JSPS KAKENHI Grant Number 17K14301.


\end{document}